\documentclass[referee]{aa}     % LaTeX A&A  Standard Fonts
\usepackage{color,graphicx,epsfig}
\usepackage{float}

\begin{document}
\thesaurus{11.03.1         % Galaxies:clusters
           11.03.3,        % Galaxies: cooling flows
           13.25.3,        % X-Rays: galaxies
           02.13.5         % Molecular processes
}
%\titlerunning{Thermal equilibrium}
\titlerunning{Thermal Equilibrium...}
\title{Thermal Equilibrium of Molecular Clouds in Cooling Flow Clusters}
\author{Denis Puy, Lukas Grenacher, Philippe Jetzer}
\authorrunning{Puy, Grenacher, Jetzer}
\offprints{puy@physik.unizh.ch}
\institute{
Paul Scherrer Institute\\ 
Laboratory for Astrophysics\\
CH-5232 Villigen PSI (Switzerland)
\\
and
\\
Institute of Theoretical Physics\\
University of Z\"urich\\
Winterthurerstrasse, 190\\
CH-8057 Z\"urich (Switzerland)}
%\date{Received; accepted}
\maketitle
\begin{abstract}
In many clusters of galaxies there is evidence for cooling flows in the 
central regions. The ultimate fate of the gas which cools is still unknown. 
A possibility is that a fraction of the gas forms cold molecular clouds. 
We investigate the minimum temperature which can be reached by clouds in 
cooling flows by computing the cooling function due to $H_2$, $HD$ and 
$CO$ molecules.  As an example, we determine the minimum temperature 
achievable by clouds in the cooling flows of the Centaurus, Hydra and 
PKS 0745-191 clusters. Our results suggest that clouds can reach very low 
temperatures - less than $\sim 10$ K - which would explain the non-detection 
of $CO$ in these clusters.
\end{abstract}
%----------------------------------------------------------------------------
\section{Introduction}
%----------------------------------------------------------------------------
In several clusters of galaxies one observes a soft $X$-ray excess 
towards the central regions. This excess is interpreted as being due
 to hot intracluster gas (10$^7$-10$^8$ K) with a cooling time
  less than the Hubble time (Fabian 1994). This gas cools and falls quasi-hydrostatically
   into the center of the cluster potential well 
   (e.g. Fabian, Nulsen \& Canizares 1984, 1991; Sarazin 1988).
\\
Indeed, radial profiles of temperature and density in the intracluster gas inferred from $X$-ray observations show that cooler denser gas (10$^6$-10$^7$ K) 
is present in the central region of the cluster (Johnstone, Fabian, Edge \& Thomas 
1992). 
Typical mass accretion rates inferred in these cooling flow clusters are 
$\sim$ 100 M$_\odot$ year$^{-1}$ and even higher in some cases (Fabian 1994). Thus cooling flows in clusters of galaxies deposit
 large quantities of cool gas around the central galaxy, which is still 
continuing to grow today. Recent extreme ultra-violet (EUV) observations 
revealed the presence of important amounts of a warm EUV emitting gas in the 
cooling flow cluster A1795 (Lieu, Mittaz, Bowyer et al. 1996a, 1996b).
\\
The final evolution of the 
cool gas is not clear. It may just accumulate as cool dense clouds, indeed 
White, Fabian, Johnstone et al. (1991) have observed a large amount of cold $X$-ray
absorbing matter distributed over the inner few 100 kpc of some
clusters of galaxies and came to the conclusion that the absorbing
component is likely to be either atomic or molecular
(mainly hydrogen). Moreover, they suggested that the gas is very cold ($< 10$ K)
and that it makes up a significant fraction
of the total mass, which has cooled out of the cooling
accretion flow during the lifetime of the cluster. 
\\
Fabian, Johnstone \& Daines (1994) have postulated that dust can form in the 
cold clouds embedded in cooling flows, which would explain why the clouds are 
almost undetectable outside the $X$-ray wavelength range. However, the nature 
of the absorbing material remains uncertain. The limits imposed by 21-cm and 
$CO$ observations, suggest that it is unlikely that significant amounts of 
cold gas can remain undetectable. On the other hand Jaffe \& Bremer (1997), 
using K-band spectroscopy, have found that the inner few kpc of central cluster
 galaxies in cooling flows have strong emission in the $H_2$(1-0)S(1) line. 
This is not seen in a comparable sample of non cooling flow galaxies. Jaffe \&
 Bremer (1997) suggest that it is likely that a large mass of additional 
molecular material is in the cooling flow, producing the soft $X$-ray 
absorption. At least a fraction of the cool gas must reach the inner regions of
 the central cluster galaxies, where it could accumulate into molecular clouds 
and subsequently form stars (Cardiel, Gorgas \& Arag\'on-Salamenca 1998).
\\ 
The metallicity of a cluster seems to be correlated with the presence
of a cooling flow (Allen \& Fabian 1998).
Mushotzky \& Lowenstein (1997) showed that most of the enrichment of the
intercluster medium occured at high redshifts. This is consistent with
the 
semi-analytic models of galaxy formation (Kauffmann \& Charlot 1998),
which find indeed
that the metal enrichment occurs at $z>1$. In this context molecules such 
as $CO$ can subsequently be formed in the gaseous medium. The molecular gas can cool
down to very low temperatures within the intra-cluster environment as
has been suggested by Ferland, Fabian \& Johnstone (1994). 
\\
O'Dea, Baum, Maloney et al. (1994) searched for molecular gas, 
by looking for $CO$ emission lines in a heterogeneous sample of five
radio-loud galaxies (three of which are in cooling flow clusters), 
using the Swedish-ESO Submillimeter Telescope in 
the frequency ranges of 80-116 GHz and 220-260 GHz. 
A positive detection of $CO$ has been made in two of the galaxies: PKS 0634-206, a classical double radio galaxy, and possibly in NGC 4696. They come to the conclusion 
that in order to have escaped detection the gas has to be very cold, close to 
the temperature of the Cosmic Background Radiation (CBR). 
O'Dea, Baum, Maloney et al. (1994) estimated also the expected temperature of the
clouds by comparing the $X$-ray heating with the cooling rate for 
molecular gas. For the latter they adopted the cooling rate calculated by
Goldsmith \& Langer (1978), which includes line cooling from both molecular
and atomic species likely to be abundant in molecular clouds. 
They thus assumed that the clouds have a metallicity similar to solar
neighbourhood clouds. However, the Goldsmith \& Langer cooling rates are
valid in the range $\sim$ 20 - 60 K, 
below $\sim$ 20 K they are no longer reliable. In addition the possible
role of $HD$ as a cooling agent was not taken into account.
Therefore, the conclusion of O'Dea, Baum, Maloney et al. (1994), that the  minimum temperature of the clouds is in the range 20-30 K is certainly questionable. 
\\
Recently, Bridges \& Irwin (1998) have observed molecular gas in the Perseus cooling flow 
galaxy NGC 1275 and have mapped the central arcminute of this galaxy with the James Clerk 
Maxwell Telescope in $^{12}CO(2-1)$. 
Assuming a galactic conversion between $CO$ intensity and the molecular $H_2$ 
gas mass they inferred a total mass M$_{H_2}\sim 1.6 \times 10^{10}$ 
M$_\odot$. They argue that the reason why $CO$ has been detected in NGC 1275 
is that the radio emission heats up the molecules to a detectable temperature. 
Indeed, NGC 1275 happens to have one of the strongest radio core of any cooling 
flow central galaxy. Without the heating of the radio emission the gas would 
stay at much lower temperatures and thus escape detection.
\\
The aim of this paper is to investigate the minimum temperature achievable
  by clouds in cooling flows by improving the analysis done by O'Dea, Baum, 
  Maloney et al. 
  (1994), in particular by considering clouds made of $H_2$, $HD$ and $CO$ 
  molecules and by computing cooling functions which are more appropriate for 
  temperatures below 20 K. 
 We do not discuss the different pathways for forming $H_2$, $HD$ and $CO$ 
 molecules. $H_2$ and $HD$ can be formed at the early epoch of the recombination 
 of the hydrogen (see Lepp \& Shull 1984, Puy, Alecian, Le Bourlot et al. 1993), whereas 
 $CO$ formation can be triggered by the radiative association of $C$
  with $O$ after the first star formation. 
 \\
 The paper is organised as follows: in Sect. 2 we 
  recall the effects of $X$-ray heating on the clouds using the results of 
  Glassgold \& Langer (1973) and O'Dea, Baum, Maloney et al. (1994). In
Sect. 3 we calculate the molecular cooling due to $H_2$, $HD$ and $CO$, 
respectively, through the transition between the ground state and the 
first rotational level, since we are interested on what happens for very cold 
clouds (with temperature of order 20 K or even below). Moreover, 
also from the observational point of view the transition between the ground 
state and the first excited state gives possibly the strongest contribution 
at very low temperatures as suggested by the measurements by Jaffe \& Bremer (1997) on the 
emission in the $H_2$(1-0) S(1) line.
 We then estimate in Sect. 4 the equilibrium temperatures of the 
clouds for  the Centaurus, Hydra and PKS 0745-191 clusters, and discuss 
the observational consequences. 
\\
Some technical details of the calculations are given in two Appendices. 
%----------------------------------------------------------------------------
\section{$X$-ray heating of the clouds}
%----------------------------------------------------------------------------
The clouds are embedded in the hot intracluster gas, whose emission is 
dominated by thermal bremsstrahlung and thus the clouds are mainly heated by 
$X$-rays. The mechanisms by which soft $X$-rays heat a gas cloud have been
extensively studied by a number of authors: Habing \& Goldsmith (1971)
and Bergeron \& Souffrin (1971) for $H_I$ regions, Glassgold \& Langer
(1973) for molecular hydrogen clouds.
\\
Due to the increased number of possible charged-particle reactions with molecules,
the heating mechanism for molecules is different and more complicated as 
compared to the heating processes for atoms alone.
Glassgold \& Langer (1973) have given a survey of the physical processes for a pure
molecular hydrogen medium. While $X$-ray heating of molecular hydrogen regions
has been extensively studied, this has not been the case for other molecular
species. Glassgold \& Langer (1973) discuss in detail the energy loss processes and present a
quantitative analysis which shows which part of the incident $X$-ray
energy will go into heat. In Appendix A we recall the heating processes 
for $H_2$, $HD$ and $CO$ molecules which are of interest for us. 
\\
In order to evaluate the
bremsstrahlung flux coming from the intracluster gas, we must define a density distribution for the electrons in the gas. A classical profile is a non-singular
isothermal sphere (Jones \& Forman 1984):
$$
n_e(R) \, = \, \frac{n_o}{ 1+\Bigl ( \frac{R}{r_c} \Bigr )^2} 
 \ {\rm cm}^{-3},
$$
where $n_o\sim 10^{-3}$ cm$^{-3}$ is the central gas number density, 
$r_c$ is the core radius and $R$ is the distance from the center of the cluster. 
In cooling flow clusters the central density is somewhat higher in the range 
$10^{-3}-10^{-2}$ cm$^{-3}$ (see Stewart, Fabian, Jones \& Forman 1984). 
\\
For temperatures $T> 2$ keV the emission from the hot gas is dominated by thermal bremsstrahlung and 
the emissivity for an ion with charge $Z$ and number density $n_i$ in a gas with electron density $n_e$ is 
\begin{eqnarray}
\epsilon_\nu \, &=& \, 2.4 \times 10^{-16} Z^2 n_e n_i T^{-1/2}_{keV} \, 
{\rm exp}(-E_{keV}/T_{keV}) \overline{g}_{ff} \nonumber \\
&=& \, \epsilon^o_{\nu} Z^2 n_e n_i \ \ {\rm (in \, keV \, cm}^{-3} \, {\rm s}^{-1} \,  
{\rm keV}^{-1} \, {\rm sr}^{-1}),
\end{eqnarray}
where $\overline{g}_{ff}$ is the velocity-averaged Gaunt factor, while gas temperature $T$ and 
photon energy $E$ are given in keV. The Gaunt factor for the relevant parameter range is accurately given by 
\begin{equation}
\overline{g}_{ff} \, \sim \, 0.55 \, {\rm ln}(2.225) \,  \frac{T_{keV}}
{E_{keV}} \ .
\end{equation}
The bremsstrahlung "flux" (i.e. $4 \pi J_E$, where $J_E$ is the mean intensity at energy $E$), seen by a cloud located at a radial distance $R$ from the center of the cluster is given by O'Dea, Baum, Maloney et al. (1994) in keV cm$^{-2}$ s$^{-1}$ 
keV$^{-1}$
\begin{equation}
4\pi J_E(R) \, = \, \frac{ 1.5 \pi n_o^2 r_c^4}{(r_c^2 + R^2)^{3/2}}\, 
\epsilon_\nu ^o \
\int^{R_b}_0 dr \, \int^{\pi}_0 d\theta \, \frac{r^2{\rm sin} \theta (r_c^2 + R^2)^{3/2}}
{(r_c^2 + r^2)^2 (r^2 +R^2 -2rR{\rm cos}\theta)} \ .
\end{equation}
The factor 1.5 takes into account the presence of He and heavy elements. 
$R_b$ is the outer boundary of the cluster gas distribution. The integral 
in Eq.(3) for typical values $r_c=100$ kpc and $R_b \sim 500$ kpc with 
$0 \leq R \leq 500$ kpc varies between 1.57 and 7. A reasonable average 
value for the integral is $\pi$, which for clouds located in the inner regions
 of the cluster is an upper bound.
\\
The problem is much more complicated if we consider the elastic scattering
of electrons from $H_2$ molecules. We have various dissociation and
ionization channels which contribute to the heating (see Glassgold \&
Langer 1973). O'Dea, Baum, Maloney et al. (1994) assume that the fraction of 
primary photoelectron's energy which goes into heating is close
to half.
\\
Moreover, O'Dea, Baum, Maloney et al. (1994) consider a possible 
attenuating column density within a cloud, which they characterize by the parameter 
$\tau$. Finally, with these different approximations they obtain for
the heating rate (in erg cm$^{3}$ s$^{-1}$):
\begin{equation}
\Gamma_X(R) \, = \, \gamma_o \, \frac{ n_{H_2} n_{-3}^2 
\, \, r_{100} }{
 \Bigl [ 1+\Bigl ( \frac{R}{r_c} \Bigr )^2 \Bigr ]^{3/2}}  \, T^{\alpha}_{keV} 
\, \tau^{\delta}~,
\end{equation}
where the parameters are given in Table 1 and $n_{-3}=n_o/10^{-3}$ 
cm$^{-3}$ and $r_{100}=r_c/100$ kpc.

\begin{table}
\begin{flushleft}
\begin{tabular}{||l|l|l|l||}
\hline
\hline
$\tau$ & $\gamma_o$ & $\alpha$ & $\delta$ \\
\hline
\hline
0.01-0.5 & $1.2 \times 10^{-28}$ & 0.1 & -0.65 \\
\hline
0.5-2.5& $4.6 \times 10^{-29}$  & 1/2 & -1.1 \\
\hline
\hline
\end{tabular}
\caption[ ]{Parameters for the $X$-ray heating (see O'Dea, Baum, Maloney et al. 1994).}
\end{flushleft}
\end{table}

%----------------------------------------------------------------------------
\section{Molecular cooling}
%----------------------------------------------------------------------------
In the standard Big Bang models primordial chemistry took place around the 
epoch of recombination. At this stage the chemical species essentially were 
$H$, $H^+$, $D$, $D^+$, $He$ and $Li$. Then with the adiabatic cooling of 
the Universe due to the expansion, different routes led to molecular 
formation (Lepp \& Shull 1984, Black 1988, Puy, Alecian, Le Bourlot et al. 1993, 
Galli \& Palla 1998). 
\\ 
Heavier molecules than $H_2$ play an important role in the
cooling of the denser clouds (Lepp \& Shull 1984, Puy \& Signore 1996, 1997, 1998a, 1998b)
 via the excitation of rotational levels. 
Ferland, Fabian \& Johnstone (1994) have investigated the physical conditions
within molecular clouds, particularly the structure of gas clouds in 
cooling flow and showed that 
molecular cooling plays an important role. 
\\
We calculate the molecular cooling in the range of temperatures $\sim$ 3-120 K, 
where 120 K corresponds to the temperature of the transition between the 
ground state ($J=0$) and the first rotational level ($J=1$) for the $HD$ molecule. 
In this range, molecules such as $H_2$, $HD$ and $CO$ are very efficient for 
cooling. $HD$ and $H_2$ are the main cooling agents between 80-120 K, then 
$HD$ dominates between  $\sim 40$-80 K ($H_2$ cooling, in this case, turns 
out to be negligible). $CO$ can be the main coolant in the range of 
temperature $\sim 3$ K up to $\sim 40$ K depending on its abundance. In
 order to calculate analytically the 
molecular cooling, we only consider the transition between the ground state 
and the first rotational level, although below 120 K it is well
  possible to excite other rotational levels of $CO$,
for instance the next higher transition ($J=1 \rightarrow J=2$) corresponds to a 
temperature of $\sim$ 11 K. The reason is that we are interested on what happens with very cold clouds
(with temperature of order 20 K or less) and by considering the higher
levels the molecular cooling gets even more efficient.
Thus by taking into account only the first level we will get a lower bound
on cooling and, therefore, an upper bound on the equilibrium temperature. 
\\ 
The cooling functions for the molecules $H_2$, $HD$ and $CO$ are computed in Appendix B. 
We have taken into account radiative transfer effects following the treatment by
 Castor (1970), Goldreich \& Kwan (1974) in a similar way as done in the paper 
 of Goldsmith \& Langer (1978). Thus, the rates for absorption and stimulated emission are given entirely 
 in terms of two locally determined quantities, the source function and the escape probability $\beta_o$ which 
 depend on the molecule (see Appendix B). Finally we obtain for the 
cooling functions for $H_2$, $HD$ and $CO$ (in erg cm$^3$ s$^{-1}$):
\begin{equation}
\Lambda_{H_2} \, \sim \, 
8.61 \times 10^{-24} \, \beta^{H_2}_o n_{H_2}^2 \sqrt{T_m}
\frac{exp(-512/T_m)}
{0.69 \beta^{H_2}_o + \sqrt{T_m} n_{H_2} \Bigl [ 1+ 5 exp(-512/T_m) \Bigr ]}
\ , \end{equation}
\begin{equation}
\Lambda_{HD} \, \sim \, 
2.66 \times 10^{-21} \,  \beta^{HD}_o \eta_{HD} \,  n_{H_2}^2 \sqrt{T_m} 
\frac{ exp(-128.6/T_m)}
{1416 \beta^{HD}_o  + \sqrt{T_m} n_{H_2} \Bigl [ 1+ 3 exp(-128.6/T_m) \Bigr ]} \ ,
\end{equation}
\begin{equation}
\Lambda_{CO} \, \sim \, 
1.55 \times 10^{-22} \,  \beta^{CO}_o  \eta_{CO} \, n_{H_2}^2 \sqrt{T_m} 
\frac{ exp(-5.56/T_m)}
{2659 \beta^{CO}_o + \sqrt{T_m} n_{H_2} \Bigl [ 1+ 3 exp(-5.56/T_m) \Bigr ]} \ ,
\end{equation}
with the ratios 
$
\eta_{HD} \, = \, n_{HD}/ n_{H_2} \ {\rm and} \ 
\eta_{CO} \, = \, n_{CO} /  n_{H_2} \ .
$ 
The various quantities entering in the above equations are defined in
Appendix B.
\\
Combes \& Wiklind (1998) have detected molecular absorption lines systems with
 the 15-m Swedish-ESO Submillimetre Telescope at high redshift. In particular, they 
 have detected $CO$ absorption lines in some galaxies or quasars. For Centaurus A
they estimated the column density for $CO$ and $H_2$ to be N$_{CO}=10^{16}$
cm$^{-2}$ and N$_{H_2}=2.0 \times 10^{20}$ cm$^{-2}$, respectively. For some clusters they could only 
derive an
upper bound of about $6 \times 10^{15}$ cm$^{-2}$ for the $CO$ column density. The values for 
 the $CO$ and $H_2$ column densities can vary substantially between different clusters. 
 \\
 Moreover, we do not know 
 if the column density is due to the sum of several smaller clouds which are present along the line 
 of sight or to a unique big cloud. It is thought that a Jean's unstable and isothermal cloud fragments 
 into a small number of subclouds, that in turn repeat the process recursively at smaller scale as long as
isothermal conditions prevail. Thus it is not unrealistic to consider very
small clouds in cooling flow as a result of fragmentation (Fabian 1992). Indeed, the high pressure in a
flow implies that the Jeans mass of a very cold cloud is small (Fabian 1992). 
We expect than that the molecules are mostly produced
in the small clouds where the densities can be higher with respect to the surrounding regions and therefore also 
the cooling will be most efficient there.
\\
These small clouds are most likely embedded in bigger less dense clouds with decreasing molecular 
abundance and also higher
temperatures. However, the gas surrounding the small clouds will attenuate the incoming bremsstrahlung flux as
taken into account by the $\tau$ parameter discussed in Section 2. 
\\
A scenario with successive fragmentation of the clouds in cooling flows has been proposed by Fabian \& Nulsen (1994)
and in another context, namely for the dark matter in the disk of a galaxy by Pfenniger \& Combes (1994), whereas in 
the halo of our galaxy by De Paolis et al (1995). Pfenniger \& Combes (1994) have proposed that the cold $H_2$ clouds
have a fractal structure.
\\
We will thus in the following assume as a picture that the clouds fragment into small ones as a 
result of cooling
processes. For definiteness we will adopt the following column densities for a typical small cloud: 
$N_{CO}=10^{14}$ cm$^{-2}$, 
$N_{H_2}=2 \times 10^{18}$ cm$^{-2}$ with $n_{H_2} = 10^6$ cm$^{-3}$ (density of $H_2$ the collisional species). 
We have the following ratio $\eta_{CO} \, \sim \, \frac{N_{CO}}{N_{H_2}} \, \sim \, 5 \times 10^{-5}$. 
O'Dea, Baum, Maloney et al. 1994 instead use the value 
$\eta_{CO} = 10^{-4}$. For $HD$ we take 
$\eta_{HD} \, \sim \, \frac{N_{HD}}{N_{H_2}} \, \sim \, 7 \times 10^{-5}$, 
which corresponds to the primordial ratio (Puy, Alecian, Le Bourlot et al. 1993). 
This should possibly be an upper limit. These characteristics correspond to a
size for the cloud of $L \sim 10^{-6}$ pc
which is close to 30 AU as for instance proposed in the fractal model of Combes \& Pfenniger (1994).
\\
Using the above values we can then estimate the escape probability for a photon emitted in a transition 
in the temperature range $\sim 3$ K to 120 K due to the three considered molecules. 
The escape probabilities due to 
$HD$ and $H_2$ are close to 1 ($\beta_o^{H_2}= \beta_o^{HD} \sim 1$) in the range where $HD$ and $H_2$ are
dominant ($T_m > 40$ K). In the case of $CO$ molecule,
 which is supposed to be the main cooling agent at very low temperature, the probability, that the photons emitted 
 during the transitions $J=1 \rightarrow J=0$ are reabsorbed, is non negligible and thus the escape  
 probability $\beta_o^{CO}$ is below 1. In Table 2 we give some values of $\beta_o^{CO}$ for different 
 temperatures. Particularly at low temperatures  re-absorption has to be taken into 
 account and leads thus to a less efficient $CO$ cooling. 

\begin{table}
\begin{flushleft}
\begin{tabular}{||l|l|l|l|l||}
\hline
\hline
$T_m$ & 2.735 & 5 & 20 & 10 \\
\hline
 $\beta_o^{CO }$ & 0.47 & 0.72 & 0.89 & 0.96 \\
\hline
\hline
\end{tabular}
\caption[ ]{Escape probability $\beta_o^{CO}$ for a photon emitted in a transition 
$J=1 \rightarrow J=0$ for the $CO$ molecule, assuming a column density 
$N_{CO} = 10^{14}$ cm$^{-2}$ of the cloud.}
\end{flushleft}
\end{table}

Our calculations differ from the ones obtained by O'Dea, Baum, Maloney et al. (1994), 
since they used the cooling rates calculated by 
Goldsmith \& Langer (1978), which include line cooling from both molecular and 
atomic species likely to be abundant in disk molecular clouds. Moreover, the possible role of $HD$ as 
a cooling agent was not taken into account. Goldsmith \& Langer (1978) considered molecular hydrogen 
densities in the range 10$^2$ to $10^5$ cm$^{-3}$, kinetic temperatures between 10 K and 60 K and 
a wide variety of molecules with fractional abundances bigger than or comparable to $10^{-9}$ 
($H_2$, $CO$, $H_2O$, $O_2$, $C_2$, $N_2$ and hydrides). They find, like us, that below 40 K $CO$ is the dominant 
coolant. Nevertheless, they 
approximated the cooling rate by a power law for temperatures between 10 K and 60 K. 
Instead, we approximate the molecular cooling by considering only the transition 
between the ground state and the first rotational level. We examine the equilibrium of clouds
for temperatures below 10 K, a range  
where the power law dependence of Goldsmith \& Langer (1978) is no longer valid.

%----------------------------------------------------------------------------
\section{Results and discussions}
%----------------------------------------------------------------------------
We have seen that molecular clouds in cooling flows are heated by the 
external $X$-ray flux. At equilibrium the heating and molecular cooling
 are equal: $\Gamma_X \, = \, \Lambda_T$, where the total cooling function is 
defined as $\Lambda_T \, = \, \Lambda_{H_2} + \Lambda_{HD} + \Lambda_{CO}$, see 
Eqs. (5), (6) and (7). 
The $X$-ray heating rate (see Eq. 4) depends on the hot gas
parameters $n_{-3}$, $r_{100}$ and 
temperature $T_{keV}$, for which we adopt the values (see Table 3) of three 
clusters considered by O'Dea, Baum, Maloney et al. (1994): 
Hydra A, Centaurus and PKS 0745-191.

\begin{table}
\begin{flushleft}
\begin{tabular}{||l||l|l|l|l||l|l||}
\hline
\hline
{\bf Cluster} & $n_{-3}$ & $r_{100}$ & 
$T_{keV}$ & {\bf References} & $r_{cool}$ & {\bf References} 
\\
\hline
\hline
Hydra A & 6.5 & 1.45 & 4.5 & David et al. 1990 &
 $162^{+56}_{-68}$ & Cardiel et al. 1998 
\\
\hline
Centaurus & 9 & 2 & 2.1 & Matilsky et al. 1985 &
$87^{+7}_{-29}$ & Allen \& Fabian 1997
\\
\hline
PKS 0745-191 & 35 & 0.5 & 8.6 & Arnaud et al. 1987 &
$214^{+29}_{-25}$ & Cardiel et al. 1998
\\
\hline
\hline
\end{tabular}
\caption[ ]{Parameters for the galaxy clusters we consider in the text, $n_{-3}$ is in units of 
10$^{-3}$ cm$^{-3}$, $r_{100}=r_c/100$ where $r_c$ is in kpc and $T_{keV}$ in keV 
(see O'Dea, Baum, Maloney et al. 1994). $r_{cool}$ is the cooling radius with upper and
 lower limits.}
\end{flushleft}
\end{table}

We have seen (Eq. 4) that  $\Gamma_X$ depends on the distance 
$R$ (from the center of the cluster) at which the molecular cloud is located. Thus at the 
equilibrium temperature $T_{eq}$, where $\Gamma(R_{eq}) \, = 
\, \Lambda_T(T_{eq})$, we obtain 

\begin{equation}
 R_{eq} \, = \, r_c \, \sqrt{ \Bigl[ \frac{\gamma_o n_{-3}^2 n_{H_2}
r_{100} T_{keV}^\alpha \tau^\delta}{\Lambda_T} \Bigl]^{2/3} -1} \ .
\end{equation}

The intracluster gas is denser in the core of the cluster and, therefore, the 
radiative cooling time, $t_{cool}$, due to the emission of $X$-rays is shortest there. A cooling flow is formed when $t_{cool}$ is less than 
the age of the system $t_{sys}$, roughly $t_{sys} \sim H_o^{-1}$ where 
$H_o$ is the Hubble constant. We define the radius $r_{cool}$ where 
$t_{cool} \sim t_{sys}$, beyond this radius we do not expect a cooling flow. 
Table 3 gives the cooling radius for the three clusters we consider.

\begin{table}
\begin{flushleft}
\begin{tabular}{||l|l|l|l|l|l||}
\hline
\hline
$\tau_1$ & $\tau_2$ & $\tau_3$ & $\tau_4$ & $\tau_5$ & $\tau_6$ 
\\
\hline
\hline
0.01 & 0.25 & 0.5 & 1 & 1.5 & 2.5 
\\
\hline
\hline
\end{tabular}
\caption[ ]{Values of the attenuation factor $\tau$, which will be used in the text.}
\end{flushleft}
\end{table}

In Figs. 1, 2 and 3 we plot the curves $R_{eq}$ as a function of 
$T_{eq}$ for different 
values of the attenuation factor $\tau$ (see Table 4) for the $X$-ray heating, 
which is due to the presence of an attenuating 
column density (O'Dea, Baum, Maloney et al. 1994). 
Upper curves in the different figures correspond to low values of $\tau$, in 
which case the $X$-ray heating is more important. Thus low equilibrium 
temperatures are achieved only at 
large distances from the cluster center.
We indicate on the plots also the cooling radius $r_{cool}$. 
We notice that the curves above $r_{cool}$ in the various figures are given just for clarity, but we do not
expect them to be realistic, since the density of the clouds will decrease in the outer regions of the cluster,
so that our adopted values for the density will be too high and thus no longer valid.
The slope of the $R_{eq}$ curves depends on the molecular cooling 
$\Lambda_T$ (see Eq. 8). For 
all cases, $H_2$ cooling turns out to be negligible for equilibrium 
temperatures below 80 K. $CO$ cooling is dominant in the range from $\sim$ 3 K up to 40 K, whereas $HD$ dominates between $\sim 40$ K and 80 K. 
%Moreover, for $n_{H_2} > 10^4$ cm$^{-3}$ we have verified that $R_{eq}$ (see Eq. 8) is almost 
%independent of $n_{H_2}$ in the range of temperatures $\sim$ 3 K - 100 K. 
\\
Fig. 1 is for the Hydra A cluster. Thermal equilibrium in the cooling flow region ($R_{eq}<r_{cool}$)
is possible for different values of $\tau$. Notice, that for 
$\tau=\tau_5$ and $\tau=\tau_6$ the equilibrium 
temperature $T_{eq}$ can be as low as $\sim 10$ K. If $ R_{eq}\sim 200$ kpc 
and $r_{cool}$ close to the upper limit (218 kpc), 
it is even possible that $T_{eq}$ is as low as $\sim 3$ K for $\tau_6$. O'Dea, Baum, 
Maloney et al. (1994) suggested, in 
order to explain the non-detection of $CO$ in this cluster, that the gas is 
molecular and very cold with temperature close to that of the CBR. Our 
calculation seems to confirm this conclusion. 
\\
Fig. 2 is for the Centaurus cluster, for which the situation 
is quite different, since the cooling radius is smaller ($\sim 87$ kpc). 
For this reason the equilibrium between 
molecular cooling and $X$-ray heating is difficult to realize at small radii. We find 
equilibrium solutions in the cooling flow region (not for $\tau_1$ however) 
for a temperature beyond $40$ K, for which $HD$ cooling is important. 
In this case the temperature of the clouds is not sufficiently low in order to 
explain the non-detection of $CO$, unless $\eta_{CO}$ is higher or the total gas mass 
is relatively small. This latter possibility has also been mentioned by 
O'Dea, Baum, Maloney et al. (1994) based on their observational upper limit for the 
$J=0 \rightarrow J=1$ transition in $CO$ for NGC 4696. 
\\
In Fig. 3  we consider the PKS 0745-191 cluster. Although the cooling radius $r_{cool} \simeq 214$ kpc 
is the largest one, the temperature  $T_{keV} = 8.6$ keV of the cluster is very high and thus the 
$X$-ray heating is very important. Therefore, for this reason the equilibrium between molecular cooling
 and $X$-ray heating for low temperatures would be possible only at large distances.
  In the region of cooling flow ($R_{eq} < r_{cool}$), we find that the 
  equilibrium temperature is possible at very low temperature 
  ($T_{eq} \sim 5$ K) for $\tau_5$, $\tau_6$ and at $T_{eq} \sim 3$ K for 
  $\tau_6$ only. Thus, as for the Hydra A cluster, this could explain 
  the non-detection of $CO$.
\begin{figure}[h]
\begin{center}
\epsfig{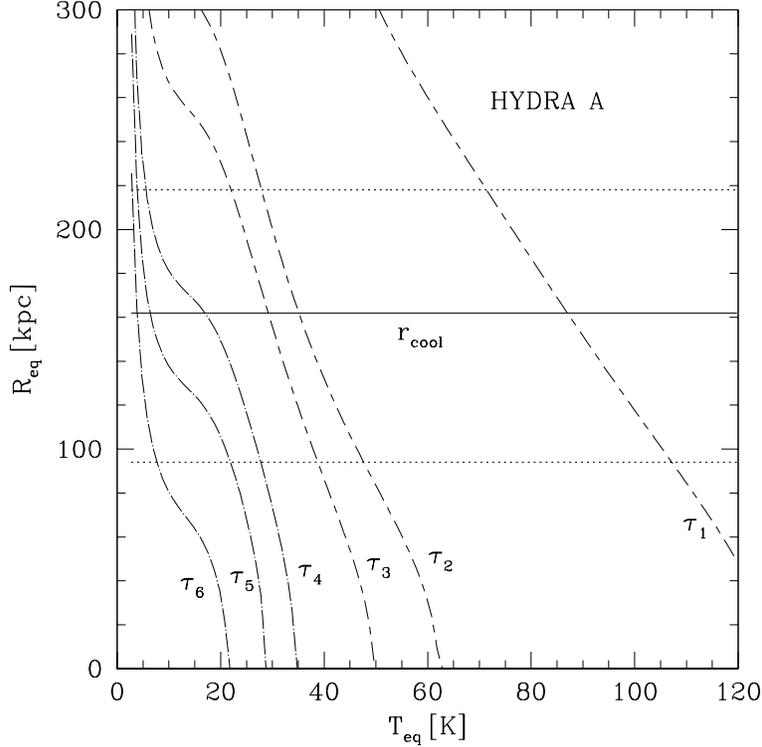}
\caption{Radius $R_{eq}$ in kpc as a function of the cloud temperature 
$T_{eq}$ for the Hydra cluster. Upper curves correspond to
low $\tau$ values. The flat curves indicate the cooling 
radius (mean value with upper and lower limit), $\eta_{HD}= 7 \times 10^{-5}$, 
$\eta_{CO}= 5 \times 10^{-5}$ and $n_{H2}= 10^{6}$ cm$^{-3}$. For $\tau_1$, $\tau_2$
 and $\tau_3$ we take for $\gamma_o$, $\alpha$ and $\delta$ the values given in the first 
 line of Table 1, whereas for $\tau_4$, $\tau_5$ and $\tau_6$ the corresponding values 
 are given in the second line.}
\end{center}
\end{figure}

\begin{figure}[h]
\begin{center}
\epsfig{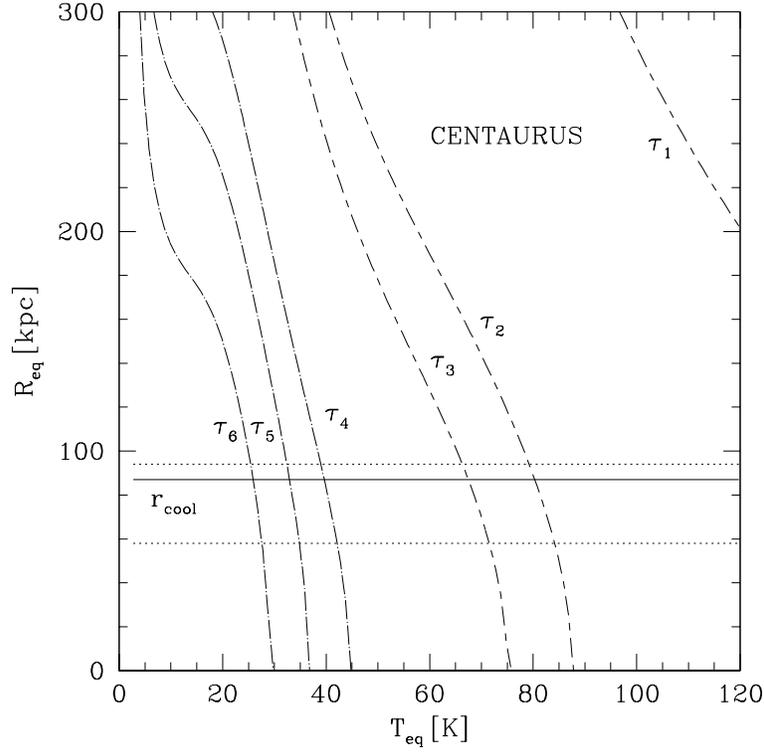}
\caption{Radius $R_{eq}$ in kpc as a function of the cloud temperature 
$T_{eq}$ for the Centaurus cluster. See also the caption in Fig. 1.}
\end{center}
\end{figure}

\begin{figure}[h]
\begin{center}
\epsfig{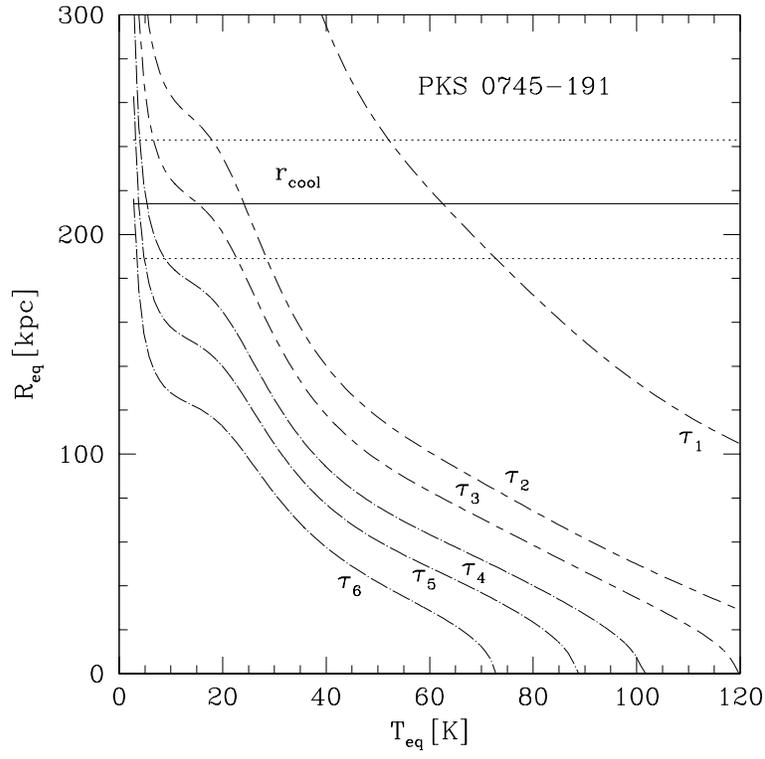}
\caption{Radius $R_{eq}$ in kpc as a function of the cloud temperature 
$T_{eq}$ for the PKS 0745-191 cluster. See also the caption in Fig. 1.}
\end{center}
\end{figure}

Of course it is difficult to draw firm conclusions on the existence of very cold clouds in 
cooling flows, since an important parameter like the $CO$ abundance is quite uncertain. In 
Figs 4, 5 we plotted the curves $R_{eq}$ (for Hydra A and PKS 0745-191) as a function of 
$T_{eq}$ for different values of $\eta_{CO}$ (see Table 5) keeping $n_{H_2}=10^6$
cm$^{-3}$, $N_{H_2}=2 \times 10^{18}$ cm$^{-2}$ and 
$\tau=\tau_6$ fixed (keeping thus the same size $L \sim 10^{-6}$ pc of the cloud and 
changing $N_{CO}$). From the figures we see that at very low temperatures $R_{eq}$ 
varies by about a factor of 2 when increasing $\eta_{CO}$ 10 times. If we took into account all transitions, in particular for 
$T_m >10$ K, the dependence on the $CO$ abundance would turn out to be
less important. We notice that there are other values of $\eta_{CO}$ for which 
it is possible to have an equilibrium temperature close to the CBR one. 
If $N_{CO}$ is substantially smaller than $10^{14}$ cm$^{-2}$ ($\eta_{CO} << \eta_{CO,1}$) 
the clouds are optically thin but cooling gets less efficient. The net effect will be an 
increase for the value of $R_{eq}$. On the other hand, if $N_{CO}$ is bigger (i.e. $\eta_{CO} > \eta_{CO,5}$) 
we have only a small decrease of $R_{eq}$ at $T_{eq} \sim 3$ K.
\\
The hypothesis of very cold molecular gas in cooling flows seems reasonable given also the fact 
that with our approximations we get upper limits for the cloud temperatures. 
We notice, that two clusters have an important cooling flow: 
400 M$_\odot$ yr$^{-1}$ for the Hydra cluster (David, Arnaud, Forman \& Jones 1990) and 
500 M$_\odot$ yr$^{-1}$ for the PKS 0745 cluster (Arnaud, Johnstone, Fabian et al. 1987), 
whereas the Centaurus cluster has a smaller cooling flow of 20 M$_\odot$ yr$^{-1}$ (Matilsky, Jones 
\& 1985). An important cooling flow could trigger thermal instabilities in the gas 
(David, Bregman \& Seab 1988,
 White \& Sarazin 1987) and could lead to the fast-growing of perturbations, which can then form objects such 
 as molecular clouds. 
\\
Other molecules than 
$CO$ could also be important, for example $CN$ or $H_2CO$ molecules. For 
$CN$ the corresponding temperature for the transition 
$J=0 \, \rightarrow \, J=1$ is $\sim 5.44$ K and for $H_2CO$ the transition is 
close to $6.84$ K (see Partridge 1995), so both molecules could play a role 
in cooling. The study of the chemistry in cooling flows could lead to important 
insight in the phenomenon and be of relevance in order to correctly 
estimate the total involved gas mass. Such a detailed 
study, however, is beyond the scope of the present paper.

\begin{table}
\begin{flushleft}
\begin{tabular}{||l|l|l|l|l|l||}
\hline
\hline
$\eta_{CO,1}$ & $\eta_{CO,2}$ & $\eta_{CO,3}$ & $\eta_{CO,4}$ & $\eta_{CO,5}$ 
\\
\hline
\hline
$10^{-5}$ & $2.5 \times 10^{-5}$ & $5\times 10^{-5}$ & $7.5 \times 10^{-5}$ & $10^{-4}$
\\
\hline
\hline
\end{tabular}
\caption[ ]{Values of the ratio $\eta_{CO}$, used in the text.}
\end{flushleft}
\end{table}

\clearpage

\begin{figure}[H]
\begin{center}
\epsfig{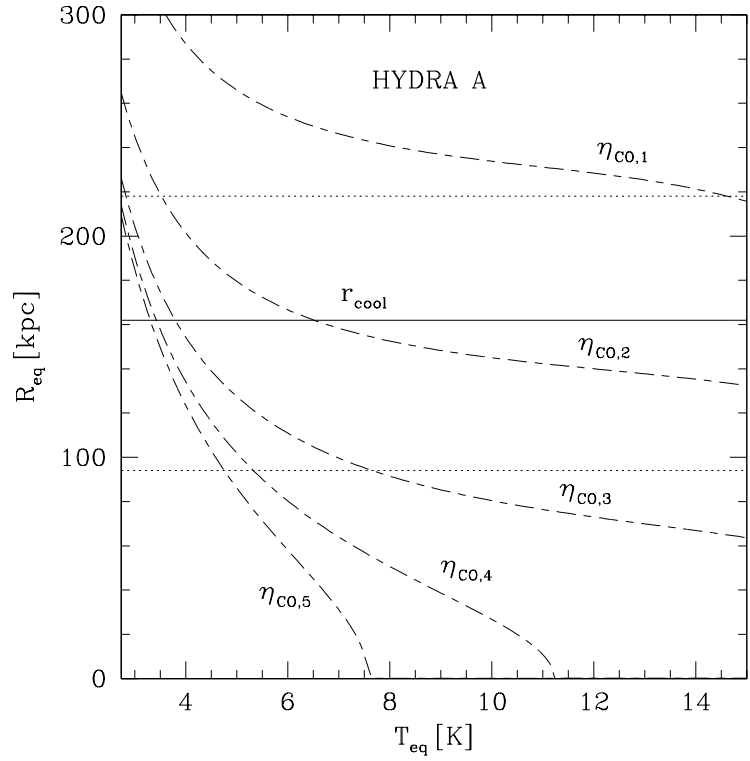}
\caption{Radius $R_{eq}$ in kpc for the Hydra cluster for different 
values of $\eta_{CO}$ ($\tau=\tau_6$, $N_{H_2}=2 \times 10^{18}$ cm$^{-2}$ and 
$n_{H_2}=10^6$ cm$^{-3}$).}
\end{center}
\end{figure}
\clearpage

\begin{figure}[H]
\begin{center}
\epsfig{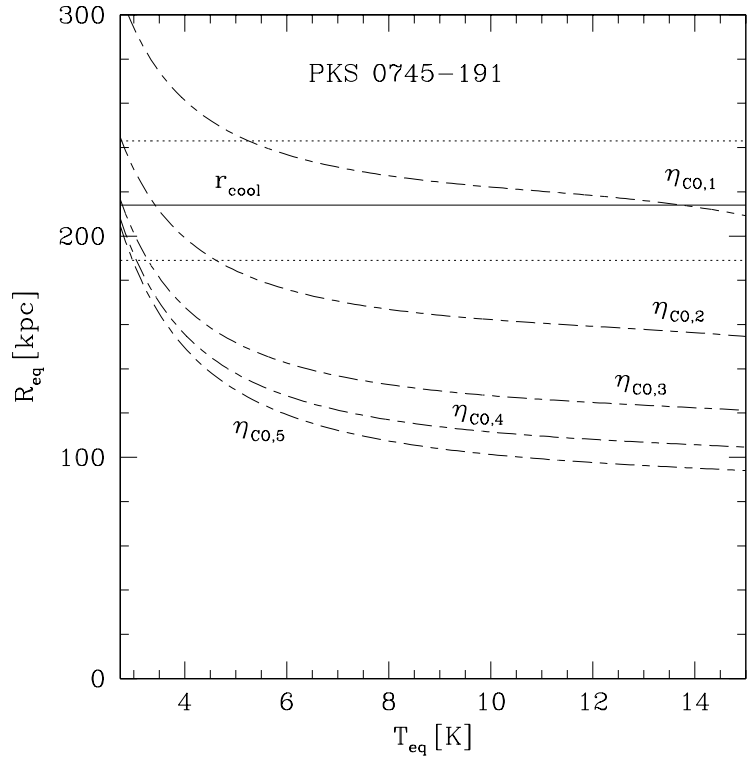}
\caption{Radius $R_{eq}$ in kpc for the PKS 0745-191 cluster for different 
values of $\eta_{CO}$ ($\tau=\tau_6$, $N_{H_2}=2 \times 10^{18}$ cm$^{-2}$ and 
$n_{H_2}=10^6$ cm$^{-3}$).}
\end{center}
\end{figure}

\subsection*{Acknowledgments}
We thank Monique Signore and Markus Str\"assle for valuable discussions. We 
are 
also grateful to the referee for his comments. This work has been 
supported by the {\it D$^r$ Tomalla Foundation} and by the Swiss National Science Foundation. 
\clearpage
%-----------------------------------------------------------------

\appendix
\section{Molecular heating}
%-----------------------------------------------------------------
In this appendix we briefly recall the heating processes for $H_2$, $HD$ and $CO$ 
molecules, which occur mainly through electronic excitations.
\subsection*{$H_2$ molecules} 
The  photoionization process of $H_2$ creates an energetic electron and an 
$H_2^+$ ion (Wishart 1979):
\begin{equation}
X_{ray} + H_2 \, \longmapsto \, H_2^+ + e^- \ .
\end{equation}
Different routes are then possible: one of the ways is that $H_2^+$ recombines 
with a thermal electron (Nakashima, Takanagi \& Nakamura 1987):
\begin{equation}
H_2^+ + e^- \, \longmapsto \, H_2 + h \nu,
\end{equation}
producing a photon but no heating, since the photon can escape from the 
cloud . The second route is that $H_2^+$ collides with $H_2$ and gives rise to the
reaction (Prasad \& Huntress 1980)
\begin{equation}
H_2^+ + H_2 \, \longmapsto \, H_3^+ + H \ ,
\end{equation}
followed by the reaction 
\begin{equation}
H_3^+ + e^- \, \longmapsto \, H_2 + H \ .
\end{equation}
These two reactions are exothermal and lead, therefore, to a transfer of energy
and a heating of 10.96 eV for every $H_2$ involved in the process (Glassgold
\& Langer 1973).
\subsection*{$HD$ molecules}
The fact that the $HD$ molecule has a permanent dipole moment 
($\mu_{HD}=8.3 \times 10^{-4}$ Debye, see 
Abgrall, Roueff \& Viala 1982) leads to $X$-ray heating processes, which are different from 
those for $H_2$. Thus, after the classical photoionization process 
(Von Bush \& Dunn 1972):
\begin{equation}
X_{ray} + HD \, \longmapsto \, HD^+ + e^-,
\end{equation}
one could expect a chemistry similar to reactions (A3) and (A4) with an intermediate species $H_2D^+$, 
that takes over the role of the $H_3^+$ ion. The knowledge of the $H_2D^+$ chemistry is, however, 
very poor and we will not discuss it in more details. The most exothermical reaction is 
\begin{equation}
HD^+ + e^- \, \longmapsto \, H + D,
\end{equation}
which gives 10.93 eV for every $HD^+$ (see Nakashima, Takanagi \& Nakamura 1987).
\subsection*{$CO$ molecules}
$CO$ has a permanent dipole moment ($\mu_{CO} = 0.118$ Debye, see Lang 1974). Exothermal chemistry relies
 on the ion $HCO^+$ (Prasad \& Huntress 1980, Anicich \& Huntress 1986). After 
 photoionisation which produces $CO^+$:
\begin{equation}
X_{ray} + CO \, \longmapsto \, CO^+ + e^-,
\end{equation}
further collision with $H_2$ molecules can produce the intermediate species
$HCO^+$:
\begin{equation}
CO^+ + H_2 \, \longmapsto \, HCO^+ + H.
\end{equation}
Afterwards the dissociative recombination
\begin{equation}
HCO^+ + e^- \longmapsto \, CO + H
\end{equation}
gives 7 eV for each process, which goes into heating.

\subsection*{Electrons}
For the electrons the situation is different. The energetic
primary electrons lose their energy by electronic excitation and ionisation,
the latter process leading to secondary electrons:
\begin{equation}
e^- + H_2 \, \longmapsto \, e_1^- + e_2^- + H_2^+ \ .
\end{equation}
Whenever the kinetic energy of a primary or a secondary-generation
electron falls below the threshold for inelastic scattering, it can
lose energy only by elastic scattering on thermal electrons and thus 
most of his energy goes into heating.
\clearpage

%-----------------------------------------------------------------
%\appendix
\section{Molecular cooling}
%-----------------------------------------------------------------

In this Appendix we compute the cooling functions for the various molecules.

\subsection*{$HD$ molecules}

The excitation of the ground state is radiative or collisional. The
equations of evolution for the populations ($X_o^{HD}$ for the
ground state $J=0$, $X_1^{HD}$ for the first excited state $J=1$) 
are given by:

\begin{equation}
\frac{dX_1^{HD}}{dt} \,= \, 
\Bigl[ C_{0,1}^{HD} + B_{0,1}^{HD} u^{HD}_{0,1} 
\Bigr ] X_o^{HD} - 
\Bigl [ A_{1,0}^{HD} +C_{1,0}^{HD} + B_{1,0}^{HD} u^{HD}_{1,0}
\Bigr ] X_1^{HD} \nonumber
\end{equation}

\begin{equation}
\frac{dX_o^{HD}}{dt} \,= \,  
\Bigl [ A_{1,0}^{HD} +C_{1,0}^{HD} + B_{1,0}^{HD} u^{HD}_{1,0}
\Bigr ] X_1^{HD} -
\Bigl [ C_{0,1}^{HD} + B_{0,1}^{HD} u^{HD}_{0,1} 
\Bigr ] X_o^{HD}, \nonumber 
\end{equation}
where $C_{0,1}^{HD}$ and $C_{1,0}^{HD}$ are the collisional
coefficients, which characterize the collisions between $HD$ and $H_2$. 
In the approximation of a Maxwell-Boltzmann distribution we have:
$$
C_{0.1}^{HD} \, = \, 3 C_{1,0}^{HD} exp\Bigl ( -\frac{T_{1,0}^{HD}}{T_m}
\Bigr )
$$
with $T_{1,0}^{HD}=T_{0,1}^{HD}=128.6$ K being the transition temperature between the states $J=0$ and $J=1$, $T_m$ the matter temperature of the cloud, $B^{HD}_{0,1}$ and
$B^{HD}_{1,0}$ the Einstein coefficients. The energy density of 
the cosmic background radiation ($T_r=2.735$ K) at the frequency $\nu_{1,0}^{HD}$ of the transition 
$J=0 \rightarrow J=1$ is
$$
u^{HD}_{1,0} = u^{HD}_{0,1} \, = \,
\frac{8 \pi h (\nu_{10}^{HD})^3}{c^3} \, \frac{1}{exp \Bigl (\frac{T_{1,0}^{HD}}{T_r}
\Bigr ) - 1} \ .
$$
Moreover, 
$$ 
A_{1,0}^{HD} \, = \, \frac{512 \, \pi^4 \, k_B^3}{3 h^4
c^3}\, \frac{B_{HD}^3 
\mu_{HD}^2}{3}
$$
is the Einstein coefficient for the spontaneous emission (see Kutner
1984) and $B_{HD} \, = \, 64.3$ K is the rotational constant (see Herzberg
1950). 
\\
We have the following normalisation for the populations: 
$X_o^{HD} + X_1^{HD} \, = \, 1$. We consider a
quasi-instantaneous transition, therefore, we can set  
$$
\frac{d X_o^{HD}}{dt} \, = \, \frac{d X_1^{HD}}{dt} \, = \, 0.
$$
With these approximations we find for the populations $X_1^{HD}$
and $X_o^{HD}$:
\begin{equation}
X_1^{HD} \, = \, 
\frac{B_{0,1}^{HD} u_{0,1}^{HD} + C_{0,1}^{HD}}
{A_{1,0}^{HD} + u_{1,0}^{HD} ( B_{0,1}^{HD} + B_{1,0}^{HD} )
+ C_{0,1}^{HD} + C_{1,0}^{HD} }
\end{equation}
and
\begin{equation}
X_o^{HD} \, = \, 
\frac{A_{1,0}^{HD} + B_{1,0}^{HD} u_{1,0}^{HD} + C_{1,0}^{HD}}
{A_{1,0}^{HD} + u_{1,0}^{HD} ( B_{0.1}^{HD} + B_{1,0}^{HD} )
+ C_{0,1}^{HD} + C_{1,0}^{HD} } \ .
\end{equation}
The probability for radiative de-excitation is
\begin{equation}
P_r^{HD} \, = \, \frac{A_{1,0}^{HD}+B_{1,0}^{HD} u_{1,0}^{HD}}
{C_{1,0}^{HD} + A_{1,0}^{HD} + B_{1,0}^{HD} u_{1,0}^{HD}} \ .
\end{equation}
The molecular cooling function 
(collisional excitation
followed by a radiative de-excitation) is given by
\begin{equation}
\Lambda_{HD} \, = \, n_{HD} X_o^{HD} C_{0.1}^{HD} P_r^{HD}
E_{1,0}^{HD},
\end{equation}
where $E_{1,0}^{HD} \, = \, k_B T_{1,0}^{HD}$ is the energy for the
transition $J=0 \, \rightarrow \, J=1$.
\\
With the following relation between the Einstein
coefficients:
$$
B_{0,1}^{HD} \, = \, 3 B_{1,0}^{HD} \ \ {\rm and} \ \ 
B_{1,0}^{HD} \, = \, \frac{c^3}{8 \pi h (\nu_{1,0}^{HD})^3} A_{1,0}^{HD},
$$
we get for the $HD$ cooling function
\begin{equation}
\Lambda_{HD} \, = \,
\frac{3 C_{1,0}^{HD} exp\bigl( -\frac{T_{1,0}^{HD}}{T_m} \bigr)
exp\bigl( \frac{T_{1,0}^{HD}}{T_r} \bigr) h \nu_{1,0}^{HD} n_{HD} A_{1,0}^{HD}}
{A_{1,0}^{HD}\Bigl [ 3+exp \bigl( \frac{T_{1,0}^{HD}}{T_r} \bigr) \Bigr ] +
C_{1,0}^{HD} \Bigl [ 1 + 3 exp \bigl( -\frac{T_{1,0}^{HD}}{T_m} \bigr) \Bigr ]
\Bigl [exp \bigl( \frac{T_{1,0}^{HD}}{T_r} \bigr) -1 \Bigr ]} \ .
\end{equation}
Since $T_{10}^{HD} >> T_r$ we can use the following approximations:
$$
3 +exp \Bigl (\frac{T_{1,0}^{HD}}{T_r} \Bigr ) 
\, \sim \, exp \Bigl ( \frac{T_{1,0}^{HD}}{T_r} \Bigr ) 
$$ 
and
$$
exp \Bigl ( \frac{T_{1,0}^{HD}}{T_r} \Bigr ) 
- 1 \, \sim \, exp \Bigl ( \frac{T_{1,0}^{HD}}{T_r} \Bigr ). 
$$
In this way we obtain for the $HD$ cooling
function:
\begin{equation}
\Lambda_{HD} \, \sim \,
\frac{3 C_{1,0}^{HD} exp(-\frac{T_{1,0}^{HD}}{T_m})
 h \nu_{1,0}^{HD} n_{HD} A_{1,0}^{HD}}
{A_{1,0}^{HD} +
C_{1,0}^{HD} \Bigl [ 1 + 3 exp\bigl (-\frac{T_{1,0}^{HD}}{T_m}\bigr) \Bigr ]} \ .
\end{equation} 
This result is valid for the optically thin limit. In the optically thick case
 one has to take into account radiative transfer effects. Goldreich \& Kwan 
(1974) computed the average escape probability for a photon
emitted in a radiative transition from the $J+1$ to the $J$ rotational level, 
denoted by $\beta_{J}$, using Castor's result (Castor 1970):
\begin{equation}
\beta_{J} = \Bigl[ 1- {\rm exp}(-\tau_{J})\Bigr]/\tau_{J}
\end{equation}
where $\tau_{J}$ is the optical depth for the transition $J \rightarrow 
J+1$. In our case we consider only the transition between the ground state
and the first excited level, and thus we get
\begin{equation}
\tau_o^{HD} \, = \, \frac{8 \pi^3 (\mu_o^{HD})^2}{3h\Delta v} N_{HD} X_1^{HD}
\Bigl[ {\rm exp}\Bigl(\frac{T_{10}^{HD}}{T_m}\Bigr) -1 \Bigr]
\end{equation}
and the corresponding average escape probability is
\begin{equation}
\beta_o^{HD} \, = \, \Bigl[ 1- {\rm exp}(-\tau_o^{HD})\Bigr]/\tau_{o^{HD}} \ .
\end{equation}
$N_{HD}$ is the column density of $HD$. The linewidth $\Delta v$ is 
assumed to be given by thermal doppler broadening 
\begin{equation}
\Delta v = \sqrt{\frac{3 k T_m}{2 m_H}} \ .
\end{equation}
 We neglect the other possible sources of 
broadening such as turbulence or large-scale systematic motions. 
\\
The $\beta_o^{HD}$ factor enters only in the spontaneous emission term. 
Therefore, the cooling function in Eq. (B8) gets modified as follows
\begin{equation}
\Lambda_{HD} \, \sim \,
\frac{3 C_{1,0}^{HD} exp(-\frac{T_{1,0}^{HD}}{T_m})
 h \nu_{1,0}^{HD} n_{HD} A_{1,0}^{HD} \beta_o^{HD}}
{A_{1,0}^{HD} \beta_o^{HD}+
C_{1,0}^{HD} \Bigl [ 1 + 3 exp\bigl (-\frac{T_{1,0}^{HD}}{T_m}\bigr) \Bigr ]} \ .
\end{equation} 
\subsection*{$CO$ molecules} The $CO$ molecule is like the $HD$
molecule and has a permanent dipole moment  ($\mu_{CO} \, = \, 0.118$ Debye)
which is much larger than the dipole moment of the $HD$ molecule (Lang 1974).
The rotational constant is low, $B_{CO} \, = \, 2.78$ K, leading to a very low
 temperature for the first transition $T_{1,0}^{CO} \, = \, 5.56$ K.  
\\
The expression for the cooling function of $CO$ molecule is obtained in the same way 
as for $HD$, and we get 
\begin{equation}
\Lambda_{CO} \, = \,
\frac{3 C_{1,0}^{CO} exp \bigl (-\frac{T_{1,0}^{CO}}{T_m} \bigr )
exp \bigl (\frac{T_{1,0}^{CO}}{T_r} \bigr ) h \nu_{1,0}^{CO} n_{CO} A_{1,0}^{CO}
\beta_o^{CO}}
{A_{1,0}^{CO}  \beta_o^{CO} \Bigl [ 3+exp(\frac{T_{1,0}^{CO}}{T_r})\Bigr ] +
C_{1,0}^{CO} \Bigl [ 1 + 3 exp\bigl (-\frac{T_{1,0}^{CO}}{T_m}\bigr ) \Bigr ]
\Bigl [exp\bigl (\frac{T_{1,0}^{CO}}{T_r}\bigr ) -1 \Bigr ]} \ ,
\end{equation}
and $\beta_o^{CO}$ si defined accordingly as in Eqs. (B9) and (B10).
\subsection*{$H_2$ molecules}
The $H_2$ molecule differs from $HD$, since it has no dipole moment and the transitions are
quadrupolar. We restrict
 our calculations to the first transition 
$J=0 \, \rightarrow \, J=2$. The other 
transitions can be neglected, since we consider matter temperatures 
which are much lower than the transition temperature of $H_2$ molecules
(512 K for the first transition). The 
Einstein coefficients for the spontaneous emission and for the first
transition are given by (see Kutner 1984):
$
A_{2,0}^{H_2} \, = \, 2.44 \times 10^{-11} \ {\rm s}
$.
\\
The relations between the Einstein coefficients are
$$
B_{0,2}^{H_2} \, = \, 5 B_{2,0}^{H_2} \ \ {\rm and} \ \ 
B_{2,0}^{H_2} \, = \, \frac{c^3}{8 \pi (\nu_{2,0}^{H_2})^3}A_{2,0}^{H_2} \ .
$$
The density of the cosmic background radiation at the transition is:
$$
u_{2,0}^{H_2} \, = \, u_{0,2}^{H_2} \, = \, \frac{8 \pi h
 (\nu_{2,0}^{HD})^3}{c^3} \, \frac{1}{exp \Bigl (
\frac{T_{0,2}^{H_2}}{T_r}-1 \Bigr ) } \ .
$$
The collisional coefficients, in the Maxwell-Bolztmann
distribution approximation, are 
$$
5 C_{2,0}^{H_2} \, = \, C_{0,2}^{H_2} exp \Bigl (
\frac{T_{2,0}^{H_2}}{T_m}
\Bigr )
$$
by analogy with the $HD$ molecule. Because $T_{2,0}^{H_2} >> T_r$ we 
can make the same approximations as for the $HD$ molecule, and 
we find for the populations $X_o^{H_2}$ and $X_1^{H_2}$:
$$
X_o^{H_2} \, \sim \, 
\frac{A_{2,0}^{H_2} +C_{2,0}^{H_2}}{A_{2,0}^{H_2} +C_{2,0}^{H_2}
\Bigl [1 + exp \Bigl ( \frac{T_{2,0}^{H_2}}{T_m} \Bigr ) \Bigr ]
} \ ,
$$
$$
X_{1}^{H_2} \, \sim \, 
\frac{
5C_{2,0}^{H_2} exp \Bigl( \frac{T_{2,0}^{H_2}}{T_m} \Bigr)}
{A_{2,0}^{H_2} +C_{2,0}^{H_2}
\Bigl [1 + exp \Bigl( \frac{T_{2,0}^{H_2}}{T_m} \Bigr) \Bigr ]
} \ .
$$
The probability of radiative de-excitation 
becomes
\begin{equation}
P_r^{H_2} \, \sim \, \frac{A_{2,0}^{H_2}}{A_{2,0}^{H_2}+C_{2,0}^{H_2}} \ ,
\end{equation}
while the molecular cooling function for the $H_2$ molecule is 
\begin{equation}
\Lambda_{H_2} \, \sim \, 
\frac{
5 n_{H_2} C_{2,0}^{H_2} \beta_o^{H_2} A_{2,0}^{H_2} h \nu _{2,0}^{H_2} 
exp\bigl( -\frac{T_{2,0}^{H_2}}{T_m}\bigr)
}
{A_{2,0}^{H_2} \beta_o^{H_2} +C_{2,0}^{H_2} \Bigl [
1+5 exp\bigl( -\frac{T_{2,0}^{H_2}}{T_m}\bigr) \Bigr ]
} \ ,
\end{equation}
again $\beta_o^{H_2}$ is defined accordingly as in Eqs. (B9) and (B10).
\subsection{Collision rates}
Since $H_2$ is the most abundant species, the collisional coefficients are 
given by 
\begin{equation}
C_{1,0}^{HD} \, = \, 
C_{1,0}^{CO} \, = \, 
C_{2,0}^{H_2} \, = \, < \sigma . v > \, n_{H_2} \ ,
\end{equation}
where  $\sigma \, \sim \, 1 \AA ^2$ is the collisional cross section, 
$n_{H_2}$ the number density of $H_2$ and $
v \, \sim \, \sqrt{\frac{3 k T_m}{2 m_H}}$, with $m_H$ being the hydrogen mass.
\clearpage

%----------------------------------------------------------------------------
%  REFERENCES
%----------------------------------------------------------------------------
{}
\end{document}